{
\documentclass{article}
 \relax
\setlength{\oddsidemargin}{.25in}
\setlength{\textwidth}{6in}
\setlength{\topmargin}{.25in}
\setlength{\textheight}{7.5in}
\usepackage{graphics}

\begin{document}

\title{A WORMHOLE-GENERATED PHYSICAL UNIVERSE WITH REFINED APPROXIMATION
}         
\author{A. L. Choudhury\footnote{Department of Chemistry and Physics, Elizabeth City State University,Elizabeth City, NC 27909} }      
\date{June 20, 2003}          
\maketitle

\begin{abstract}
   {  We improved the approximation of the model of the wormhole generated physical universe constructed by Choudhury and Pendharkar. We show here that the negative pressure of the unphysical wormhole can generate the right physical condition to explain the correct sign of the deacceleration parameter $q_0$. The negative sign of $q_0$ implies an accelerating universe.  }
\end{abstract}
\section{Introduction}
   {$\;$$\;$    The model constructed by Choudhury and Pendharkar [2](see also [1]) is one where our physical universe is subjected to a negative pressure originating from the wormhole occupying the core of the physical universe. It was assumed that the physical universe is separated from the wormhole by a thin  wall which keeps the contents of these two entities separate. However the pressure generated in the wormhole is assumed to satisfy the adiabatic gas law. The wormhole core is a modified Gidding-Strominger [3],[4] wormhole. The scale factor of the wormhole has been calculated in an Euclidean space. We then introduce the real time by analytic continuation assuming that such continuation is mathematically permissible.The scale factor of such a wormhole is found to have both positive and negative signs.We chose the nagative sign for the wormhole.This is of  an impenetrable and unphysical wormhole core. It is one possible way to explain some physical properties of the real world.Such mathematical devices to explain physical properties of the real world have been used before for example by the use of indefinite metric in field theories. We consider here that the wormhole core is an unphysical interior. However it can generate relevant effects in the physical universe. We find that the negative pressure is such an effect.The generation of such pressure can easily explain the accelerating universe.

      In our previous paper we have reached the conclusion that we live in an accelerating universe under certain approximation. In this paper we improve our approximation further and find our result does not change.

      In section 2 we give a short introduction of the  basics of the modified Gidding-Strominger wormhole model for the sake of completeness.In section 3, we show, following Choudhury and Pendharkar, how to calculate pressure. This pressure is generated inside the wormhole core and transferred to the physical universe.We also state the relevance of the approximation.  In section 4 we derive the ratio $\frac {dR} {dt}$/R of our physical universe. In section 5 we calculated the time dependent $H_0$ and the deacceleration parameter $q_0$. The discussion of results follows in section 6.}
\section{The wormhole core}
{$\;$$\;$ Choudhury and Pendharkar [2] have developed a model where they have generated a negative pressure from a wormhole located at the core of our physical universe. The core wormhole is a modified Gidding-Strominger one. Choudhury and Pendharkar have shown that one of the outcomes of the model is a negative pressure under certain approximation. In this paper we intend to relax the restriction further and show that the model's main outcome still holds.

$\;$$\;$As usual we start from a modified Gidding-Strominger wormhole action

\begin {equation}
S_E=\int {{d^4}x {{L_G}^c}(x)}+\int {{d^4}x {{L_{SA}}^c}(x)}=S_G+S_A,
\end{equation}
{where}
\begin {equation}
{{L_G}^c}(x)=\frac{{\surd g^c}{R(g^c)}} {2{\kappa}^2},
\end{equation}
{and}
\begin{equation}
{{L_{SA}}^c} (x)={\surd {g^c}}[{\frac1 2}(\bigtriangledown \Phi)^2+{g_p}^2 {\Phi}^2 Exp(\beta {\Phi}^2)]{H_{\mu\nu\rho}}^2.
\end{equation}
{In the above expression as well as in the subsequent formulae the suffix c stands for the core. Following Gidding and Strominger, we introduce}
\begin{equation}
H_{\mu\nu\rho}=\frac n {{g_p}^2 {a^3 (t)}} \varepsilon_{\mu\nu\rho},
\end{equation}
{ We get Eq.(3)}
\begin{equation}
H^2=\frac{6D} {{g_p}^2 {a^6}},
\end{equation}
{with}
\begin{equation}
D=\frac{n^2} {{g_p}^2}
\end{equation}
{The space-time interval in Euclidean space is given by}
\begin {equation}
ds^2=dt^2+a^2(t)(d\chi^2+sin^2 \chi d\theta^2+sin^2 \chi sin^2 \phi d\phi^2).
\end{equation}
{The variation of $g_{\mu\nu}^c$ in the core leads to the following equation}
\begin{equation}
G_{\mu\nu}^c=R_{\mu\nu}-{\frac1 2}g_{\mu\nu}^c R^c=\kappa^2[\nabla_\mu \Phi \nabla_\nu \Phi-({\frac1 2}(\nabla\Phi)^2) g_{\mu\nu}^c+{g_p}^2 Exp(\beta {\Phi}^2)({H_\mu}^\alpha\gamma H_\nu\alpha\gamma- {\frac1 6} g_{\mu\nu}^c H_{\alpha\beta\gamma} ^2)]
\end{equation}
{Using  the $\Phi$-variation inside the core we find the equation of motion of $\Phi$  }
\begin{equation}
{\nabla}^2 \Phi-2{{g_p}^2\beta}\Phi{Exp(\beta\Phi^2)}{H_{\alpha\beta\rho}}^2=0.
\end{equation}
{ The Hamiltonian constraint yields}
\begin{equation}
(\frac1 a \frac{da} {dt})^2-\frac1 {a^2}=\frac{\kappa^2} 3 [{\frac1 2}(\frac{d\Phi} {dt})^2-6Exp(\beta\Phi^2) \frac{n^2} {g_p^2 a^6}].
\end{equation}
{The dynamical equation yields }
\begin{equation}
{\frac d {dt}} ({\frac1 a }{\frac{da} {dt}})+\frac1 {a^2}=-\kappa^2[(\frac{d\Phi} {dt})^2-12 Exp(\beta\Phi^2) {\frac{n^2} {g_p^2 a^6}}].
\end{equation}
{ Introducing a new variable $\tau$ defined by the relation}
\begin{equation}
d\tau=a^{-3} dt,
\end{equation}
{the Eq.(9) reduces to}
\begin{equation}
\frac{d^2 \Phi} {d\tau^2}-2D\beta Exp(\beta\Phi^2)=0.
\end{equation}
{The Eqs.(9)- (11) yield (see [2] for details)}
\begin{equation}
{(\frac1 a }{\frac{da} {dt}})^2-a^4+{a_c}^4=0,
\end{equation}
{where}
\begin{equation}
a_c=\surd(\frac{\kappa^2C_o} 2)
\end{equation}
{and $C_o$ is a constant which appears while converting Eq.(13)into an integral equation [2].}
{$\;$$\;$The solution of the Eq.(14) has been obtained by Gidding and Strominger and is given by}
\begin{equation}
a^2(\tau)={a_c}^2\surd(sec(2{a_c}^2\tau))
\end{equation}
{The scale factor can now be of two possible forms }
\begin{equation}
a(\tau)=\pm a_c (sec(2{a_c}^2\tau))^{\frac1 4},
\end{equation}
{specified by the signs. Both the solutions stand on equal footing.}
\section{Pressure generated by the wormhole}
{$\;$$\;$ Following Choudhury and Pendharkar [2] we assume that the wormhole is in a gaseous state satisfying the adiabatic gas law    
\begin{equation}
P_w V_w ^\gamma=Constant=B_1,
\end{equation}
{where $\gamma$  is a constant. The volume of the wormhole can be shown to be proportional to $a^3 (\tau)$ . The pressure can be expressed as }
\begin{equation}
P_w={(\pm 1)^{-3\gamma}}B{a_c}^{-3\gamma}[sec(2{a_c}^2 \tau)]^{-(3\gamma/4)}
\end{equation}
{where B is a new constant.
$\;$$\;$We now switch from Euclidean space to Lorentz space. Instead of the  zeroeth approximation [2], we replaced, in Eq.(12), $a$ with $a_c$, and t with (it). In our present approximation we get}
\begin{equation}
d\tau={\pm}{a_c}^{-3}{\frac{dt}  {[cosh{\frac{2t} {a_c}}]^{3/4}}}
\end{equation}
{For large t we neglect the negative exponential in cosh term and obtain for $\tau$ the value}
\begin{equation}
\tau=-({\pm}){\frac 1 {3{a_c}^2}}Exp(-{\frac{3t} {2{a_c}}})
\end{equation}

{Substituting this value of $\tau$ Eq.(17) a($\tau$) changes into}
\begin{equation}
a(\tau)={\pm}{a_c}[sec({\frac 1 3}Exp(-{\frac{3t} {2{a_c}}} )]^{\frac1  4}
\end{equation} 

{Following Choudhury and Pendharkar [2] we can convert Eq.(19)into}

\begin{equation}
P_w={\pm }B{a_c}^{-3\gamma}[cos({\frac 2 3}Exp(-{\frac{3t} {2{a_c}}} )]^{(3\gamma/4)}.
\end{equation}
{$\;$$\;$As usual we have restricted $\gamma$ by the equation}
\begin{equation}
3\gamma=4n+1
\end{equation}
{where n is an integer. We choose in our model only the negative sign for the pressure $P_w$, i.e.,}
\begin{equation}
P_w=-B{a_c}^{-3\gamma}(cos[{\frac 2 3}Exp[-{\frac{3t} {2{a_c}}}])^{(3\gamma/4)}
\end{equation}
\section{ The Robertson-Walker space with a wormhole at its core}
{$\;$$\;$ As assumed by Choudhury and Pendharkar we live in a physical universe with a wormhole at the core.We reiterate the assumption  that there is a wall which separates the physical universe from the content of the wormhole but the pressure generated in the wormhole can be transferred to the physical universe through the wall.
$\;$$\;$ Thus the Robertson-Walker metric is given by the relation}
\begin{equation}
d {s_R}^2 = -d t^2 + {R^2}(t)[\frac{dr^2} {1-kr^2} + {r^2}d \theta^2 + {r^2}{sin^2}{\theta} d\phi^2]
\end{equation}
{where we have chosen c=1. The Einstein equation is as follows}

\begin{equation}
G_{\mu\nu}=R_{\mu\nu}-{\frac1 4}g_{\mu\nu} R = -{\frac{8 \pi G} 3}S_{\mu \nu} = - {\frac{8 \pi G} 3} (T_{\mu\nu}- {\frac1 2}g_{\mu\nu} {T^{\lambda}}_\lambda).
\end{equation}
{We assume that the energy momentum tensor has the perfect fluid form}
\begin{equation}
T_{\mu\nu}= P_T g_{\mu\nu}+(P_T +\rho) U_\mu U_\nu
\end{equation}
{Following Choudhury and Pendharkar we can straight forward deduce}
\begin{equation}
(\frac{dR} {dt})^2 + k = -{\frac {8 \pi G} 3} P_T R^2.
\end{equation}
{In this paper we only study the case of k=0. We find that}
\begin{equation}
\frac{(\frac{dR(t)} {dt})^2} {R^2(t)} = -{\frac {8 \pi G} 3} P_T .
\end{equation} 
{Assuming further that the fluid pressure of the physical world is significantly smaller than the wormhole pressure, we can replace $P_T$ by $P_W$ of Eq.(25). Therefore we can write }
\begin{equation}
\frac{(\frac{dR(t)} {dt})} {R(t)} = \pm \surd(-{\frac {8 \pi G} 3} P_W ).
\end{equation}
{where $P_W$ is given by the Eq.(25)}
\section{Hubble and deacceleration parameters}
{$\;$$\;$Following Choudhury and Pendharkar we can show that}
\begin{equation}
{\frac{\frac{dR(t)} {dt}} {R(t)}}= \pm i^{-3\gamma +1+4n} \surd( {\frac{8 \pi GB} 3}) {a_c}^{-3\gamma/2}(cos[{{\frac2 3}Exp[-\frac{2t} {a_c}]}])^{3\gamma/8} .
\end{equation}
{We choose n to be a positive integer and to satisfy the condition}
\begin{equation}
3\gamma =4n + 1.
\end {equation}

{Choosing now the positive sign in Eq.(32)we get}

\begin{equation}
{\frac{\frac{dR(t)} {dt}} {R(t)}}= \surd( {\frac{8 \pi GB} 3}) {a_c}^{-3\gamma/2}(cos[{{\frac2 3}Exp[-\frac{2t} {a_c}]}])^{3\gamma/8} .
\end{equation}
{For $t=t_o$, we get the time dependent Hubble parameter}
\begin{equation}
H_o(t_o)= \surd( {\frac{8 \pi GB} 3}) {a_c}^{-3\gamma/2}(cos[{{\frac2 3}Exp[-\frac{2{t_o}} {a_c}]}])^{3\gamma/8} .
\end{equation}
For large $t_o$ we get}
\begin{equation}
H_o(\infty)= \surd( {\frac{8 \pi GB} 3}) {a_c}^{-3\gamma/2}
\end{equation}
{ We can now easily find the deacceleration parameter.Following Choudhury and Pendharkar we can show that $q_o$ turns out to be:}
\begin{equation}
q_o(t_o)= 4{\pi} GB[ \gamma {{a_c}^{-(3 \gamma +1)}} e^{-\frac{2t_o} {a_c}} cos^{\frac{3\gamma-4} 4}({\frac2 3}e^{-\frac{2t_o} {a_c}})sin({\frac2 3}e^{-\frac{2t_o} {a_c}}){H_o}^{-3}-2{{a_c}^{-3 \gamma }}cos^{\frac{3\gamma} 4}({\frac2 3}e^{-\frac{2t_o} {a_c}}){H_o}^{-2}].
\end{equation}
{For large $t_o$, $q_o(t_o)$ becomes}
\begin{equation}
q_o(t_o)= -{\frac{8 \pi GB} 3} {{a_c}^{-3 \gamma }}cos^{\frac{3\gamma} 4}({\frac2 3}e^{-\frac{2t_o} {a_c}}){H_o}^{-2}.
\end{equation}
{This expression is always negative. As a result the universe expands at an accelerating rate for large $t_o$. For infinite $t_o$ the deacceleration parameter becomes }
\begin{equation}
q_o(\infty)=-1.
\end{equation}

\section{Concluding Remarks}
{$\;$$\;$ Using the model elaborated by Choudhury and Pendharkar and improving the approximation more rigorously, we have again reached the conclusion that the physical universe accelerates for large value of the present time $t_o$. In [2] the approximation $\tau$ connects  to the real time t by the equation}
\begin{equation}
\tau=i{a_c}^3 t.
\end{equation}
{Whereas in our present approximation we get Eq.[21]} 
\begin{equation}
\tau=\pm {\frac1 {3{a_c}^2}}{e^{-\frac{3t} {2{a_c}}}}.
\end{equation}
{With our new approximation we find that the physical universe driven by adiabatic pressure expands and accelerates. This is one way to explain the present experimental data [5] without introducing dark energies [6] in the picture.

$\;$$\;$ It would be really interesting to estimate the parameters introduced in the theory by comparing them with the experimental results. Work is in progress along this line.}
  
\section{References}
\begin{enumerate}
\item { A. L. Choudhury, Hadronic J., 23, 581 (2000).}
\item { L. Choudhury and H. Pendharkar, Hadronic J. 24, 275 (2001).}
\item { S. B. Giddings and A.  Strominger , Nucl. Phys. B 307, 854 (1988).}
\item { D. H. Coule and K. Maeda, Class. Quant. Grav. 7, 955 (1990).}
\item { N. Bahcall, J. P. Ostriker, S. Perlmutter, and P. J. Steinhardt, Science, 284,1481 (1999).}
\item { C. Aremendariz Picon, V. Mukhanov and Paul J. Steinhardt: Essentials of k-Essence. ArXiv:astro-ph/0006373 (2000).}
\end{enumerate}    


\end{document}